# White-light continuum emission from a solar flare and plage

Arkadiusz Berlicki[1,2], Arun Kumar Awasthi[1], Petr Heinzel[2] and Michal Sobotka[2]

[1]Astronomical Institute, University of Wrocław, Poland
email: berlicki@astro.uni.wroc.pl
email: awasthi@astro.uni.wroc.pl

[2]Astronomical Institute, Czech Academy of Sciences, Czech Republic
email: pheinzel@asu.cas.cz

**Abstract.** Observations of flare emissions in the optical continuum are very rare. Therefore, the analysis of such observations is useful and may contribute to our understanding of the flaring chromosphere and photosphere. We study the white light continuum emission observed during the X6.9 flare. This emission comes not only from the flare ribbons but also form the nearby plage area. The main aim of this work is to disentangle the flare and plage (facula) emission. We analyzed the spatial, spectral and temporal evolution of the flare and plage properties by analyzing multi-wavelength observations. We study the morphological correlation of the white-light continuum emission observed with different instruments. We found that some active region areas which produce the continuum emission correspond rather to plages than to the flare kernels. We showed that in some cases the continuum emission from the WL flare kernels is very similar to the continuum emission of faculae.

## 1. White-light emission from the solar atmosphere.

Enhanced white-light (WL) continuum emission was observed on the Sun for a long time. In most cases it is manifested by an intensity increase around active regions - such brighter areas correspond to faculae. They are overlaid by plages: bright areas observed in the chromospheric spectral lines. Faculae are often observed even through small solar telescopes using broad-band optical filters.

Besides of faculae, enhanced WL continuum emission is sometimes associated with flares. The first solar flare was just discovered in WL by Carrington (1859). It is much more difficult to 'catch' solar flare in the broad-band (WL) optical range so their observations obtained in this range are rare but desirable. Analysis of the continuum emission in solar flares is an important source of information about the response of the lower solar atmosphere to flare energy input. WL emission is thus fundamentally connected to the flare energy release, but its generation mechanism is still undefined: is it mostly chromospheric hydrogen free-bound recombination, as deduced from older blue/optical spectra (Neidig 1983), or is it $H^-$ radiation from a heated photosphere as inferred from superposed analysis of color photometry (Kretzschmar 2011)? How deep does the WL appears within the whole span of the solar atmosphere? WL flare emission can carry a large amount of the flare radiated energy, and is often spatially and temporally correlated with the impulsive heating by non-thermal electrons inferred from standard thick-target modeling of the hard X-ray observations (Hudson et al. 1992; Fletcher et al. 2007; Kowalski et al. 2015).

For a long time white light emission has been thought to be associated only with large events, but recently it has been showed that it can be observed also in very modest flares (Jess et al. 2008; Kowalski et al. 2015), although in a very limited spatial extension and temporal duration. Even today, it is not known how common the WLs are due to either lack of enough spatial resolution in the observations, or due to the limited number of flare observations suitable to determine the presence of the WL component.

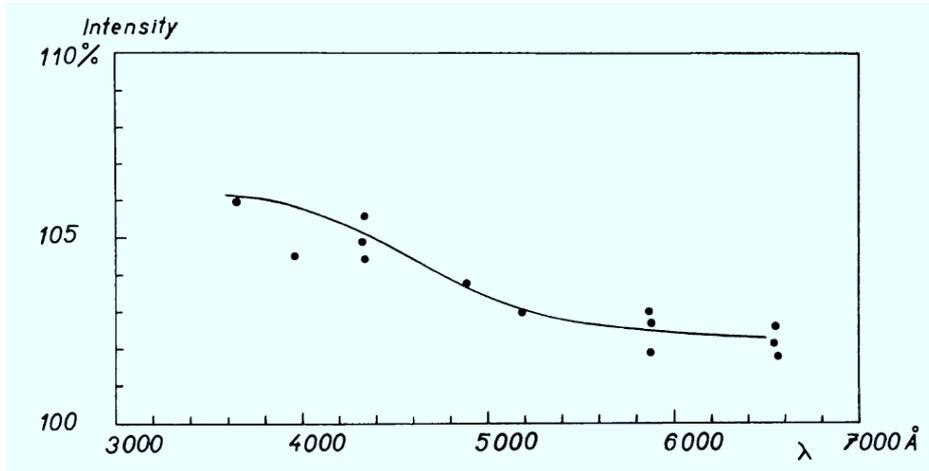

**Figure 1.** Representative example of the continuum radiation emitted from the flare observed on August 27, 1956 (according to Severny 1958).

Most of WL flare observations come from the past. Older broad-band high resolution classical spectrographs were able to record the enhanced continuum emission and Ellison (1946) was probably the first who observed the continuum spectrum of the flare. Many authors tried to determine the intensity or contrast of the WL continuum emission and they obtained the contrast values from a few to 50% (e.g. Svestka 1966; Neidig et al. 1993). Severny (1958) showed that the continuum intensity contrast in WL flare depends on the wavelengths and varies from 2 to 6% (Fig. 1). Also Machado & Rust (1974) showed that the contrast of the continuum emission depends of wavelengths and has the value from 12% at 3550 Å to 2% at 4300 Å. Recent work by Kowalski et al. (2015) confirmed a 3-8% continuum contrast for WL flare in the range 3700-4400 Å. There is a strong similarity of the time variation of the continuum emission of WL solar flares to the hard X-rays and H$\alpha$ emissions changes. This suggests that WL flares can be associated with non-thermal electrons accelerated in the corona.

As mentioned above, observations of the WL emission from faculae are more frequent. High-resolution images confirm that faculae consist of the fine scale, sometimes unresolved bright elements (Ortiz et al. 2006, and references therein). In solar continuum spectra faculae are manifested by the spatially thin brighter structures extending through the optical range - see the example in Fig. 2 .

The observed contrast of the WL continuum emission enhancement in faculae vary from a few to 10%, depending on the magnetic field strength, wavelength and the position on the solar disk (e.g. Ortiz et al. 2006). It is thus very similar to the contrast of WL emission observed for different flare kernels. It is possible that using only spectra and without detailed analysis of the time evolution of images, complemented by hard X-ray data, it is difficult to distinguish the continuum emission from flares and faculae.

Recently, we had an opportunity to analyse such a data - mixed WL continuum emission from faculae and flare. The above mentioned observations were obtained on August 9, 2011 with the Coimbra Spectroheliograph (Garcia et al. 2011).

The spectroimaging data cover a spectral range of $\pm 17$ Å around the H$\alpha$ line, thus including a quasi-continuum in the red part of the solar spectrum. The data contains the flare and faculae (plages) phenomena so their line and continuum emission can be disentangled. The observed flare was located in the active region close to the solar limb. Therefore, the intensity contrast of the WL structures was enhanced due to the limb darkening effect. Because faculae and plages are closely related, in the following sections we will use both names for this phenomenon.

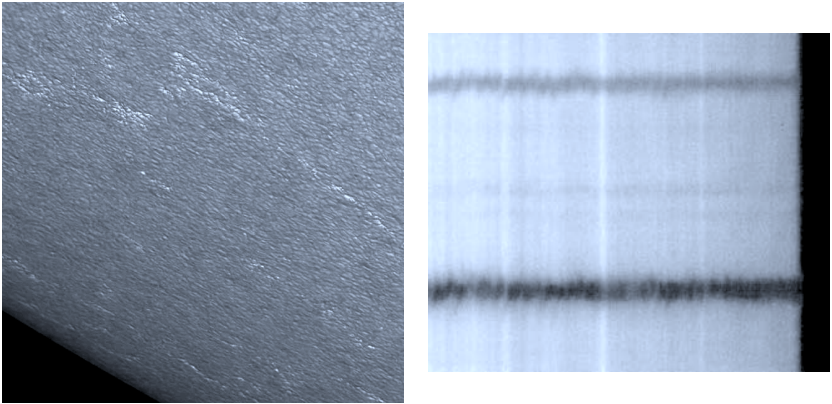

**Figure 2.** Left panel: Example of the limb facular regions in the 4875 Å continuum; field of view: approximately 80 x 80 arcsec (Hirzberger & Wiehr 2005). Right panel: Example spectrum of the limb faculae with Fe 5434 Å and Ni 5435 Å lines (Stellmacher & Wiehr 2001).

## 2. SOL2009-08-09T0808: Overview of the flare

We analyse multi-wavelength emission recorded during an X6.9 flare which occurred on August 9, 2011 in active region AR11263. AR11263 was located close to the west limb of the solar disk and had a magnetic configuration - $\beta\gamma\delta$. We make a detailed investigation of the spatial and temporal evolution of the white light continuum emission and its association with the multi-wavelength emission observed during the flare.

We use X-ray observations from Reuven Ramaty High Energy Solar Spectroscopic Imager (RHESSI; Lin et al. (2002)) mission. RHESSI observations are capable of providing temporal and spatial information of X-ray sources in 3 keV - 20 MeV energy band with the energy resolution of $\sim$1-5 keV and temporal cadence as good as 4 s. For this flare, we use the RHESSI observations during 08:00 - 08:21 UT, after which observations stopped because of RHESSI night. Further, particle emission also significantly affected the observations, however only prior to 08:00:00 UT. Next, morphological evolution of the flare plasma in the extreme ultra-violet (EUV) waveband is studied employing the observations obtained from Atmospheric Imaging Assembly (AIA; Lemen et al. 2012) on-board Solar Dynamic Observatory (SDO). SDO is a space based mission which uninterruptedly observes full-disk Sun in several extreme ultra-violet (EUV) spectral lines. Spatial and temporal resolutions of AIA/SDO observations are 0.6″per pixel and 12 seconds, respectively. The magnetograms obtained from Helioseismic Magnetic Imager (HMI; Scherrer et al. 2012) on-board SDO are also employed for this study. HMI provides the line-of-sight magnetic field measurements using Fe I absorption line at 6173 Å at the solar surface in the form of full-disk magnetograms with spatial resolution of 0.5″per pixel and temporal cadence of 45s. Apart from the magnetic field estimates, HMI instrument also provides continuum intensity map of the full solar disk which were employed for analysing the evolution of white light emission and its spatial and temporal correlation with the emission in other wavelengths.

As mentioned above, this flare was also observed by the Coimbra spectroheliograph (see Bualé et al. 2007). During regular flare patrol mode observations, this flare region was registered during 08:08:17.26 - 08:08:18.55 UT. This spectrograph provides spectral mode data in 3933.7Å (Ca II K3), 3932.3Å (Ca IIK1), 6558.7Å (H$\alpha$), and 6562.8Å (red continuum) wavelengths, while the bandwidths are 0.16 nm for Ca IIK and 0.025Å for H$\alpha$ and continuum covering spectral range of 35 Å. Spatial resolution of the observations is 2.2″/pixel (Klvaňa, Garcia & Bumba 2007).

## 3. Spatio-temporal correlation of photospheric continuum enhancement with the multi-wavelength emission

We study the spatial and temporal evolution in the photospheric continuum enhancement (hereafter PCE) as recorded by HMI/SDO. Next, the association of PCE with the loop-morphology is studied employing the co-temporal EUV images obtained from AIA/SDO. In Fig. 3, we present the relative intensity (RI) enhancement estimated from EUV images obtained in 94, 131 and 304 Å channels from AIA/SDO. The relative intensity (RI) enhancement is estimated as follows:

$$RI(\lambda, t) = \frac{I_f(\lambda, t) - I_f(\lambda, t_b)}{I_f(\lambda, t_b)} \qquad (3.1)$$

where $I_f(\lambda, t)$ is the average intensity of the selected flaring region and estimated from the image corresponding to wavelength '$\lambda$' at time 't'. Similarly, $I_f(\lambda)$ is the average background intensity of the same region and estimated from the image acquired before the commencement of the flare ($t_b$). We also overplot the relative intensity of photospheric continuum enhancement, estimated in the aforesaid manner, from the continuum images obtained by HMI/SDO instrument.

It may be noticed from Fig. 3 that the rate of relative continuum intensity enhancement is faster than that estimated from the EUV images. Moreover X-ray emission in 30-100 keV, plotted in the bottom panel of the figure, shows the evolution co-temporal with the continuum enhancement. A slow decay or the plateau region in the continuum as well as hard X-ray emission is also noticed during 08:03:00 - 08:04:25 UT, which may be attributed to continuous beam of non-thermal electrons.

Next, we study the spatial association of the region of continuum enhancement with the multi-wavelength emission during the flare. In this regard, we process the EUV images obtained from AIA/SDO. In Fig. 4, we show the EUV images during the peak of the impulsive phase of the flare (08:01 - 08:03 UT).

From the correlation of flare emission observed in EUV and HMI continuum images, as shown in Fig. 4, we may notice that the continuum emission in the form of two-ribbons is co-spatial to the foot-point location of multiple active flaring loops.

In the next step we investigate the spatial and temporal evolution of the X-ray sources in conjunction with the continuum emission. In this regard, we use RHESSI observations to synthesize images in 6-12, 12-25 and 25-50 keV energy bands with 12 sec time interval and 1 arcsec/pixel spatial resolution during 08:00 - 08:20 UT using PIXON image reconstruction algorithm (Hurford et al. 2002). Synthesized X-ray images show the correlation of the X-ray loop and foot-point emission with the continuum ribbon-shape enhancement. In Fig. 5, we overplot the contours of 3% of the maximum intensity of 6-12, 12-25 and 25-50 keV images drawn by red, blue and black colours respectively over the consecutive difference continuum filtergrams, obtained from HMI/SDO.

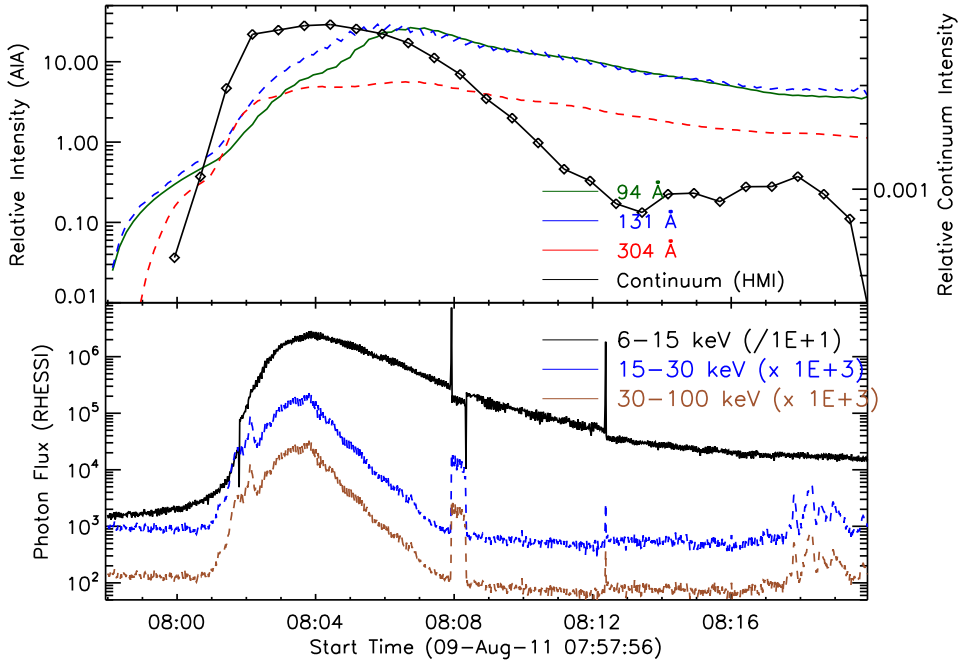

**Figure 3.** Top panel: Temporal evolution of relative intensity estimated from the images of 94, 131 and 304 Å, obtained from AIA/SDO. Continuum emission enhancement estimated from HMI observations is over-plotted with black colour. Bottom panel: X-ray photon flux in 6-15, 15-30 and 30-100 keV, derived from RHESSI observations, and plotted in the black, blue and brown colours, respectively.

It may be seen from Fig. 5 that the continuum ribbons are placed form adjacent to the apparent foot-point location of the loop. This is in agreement with the morphological correlation study made from EUV images. This result suggests continuum emission to be originated as a consequence of non-thermal electron beams bombardment.

## 4. Intensity contrast of the continuum enhancement

Although Coimbra observations were not available during the peak of impulsive phase of the flare, the spectroheliogram observations show continuum enhancement primarily in two regions R1 and R2, as shown in Fig. 6.

In this regard, we study the temporal evolution of contrast derived from the HMI continuum filtergrams. Next, we compare the HMI contrast value estimated at Coimbra observation time i.e. 08:08:17 UT, with the spectral evolution of contrast derived from Coimbra spectroheliograms. Firstly, we degrade the HMI spatial resolution to match with that of Coimbra observations. Next, we estimate the contrast of the HMI as well as Coimbra observations as per the following equation which is similar to the equation 3.1, employed for relative intensity:

$$I_c = (I_f - I_b)/I_b. \qquad (4.1)$$

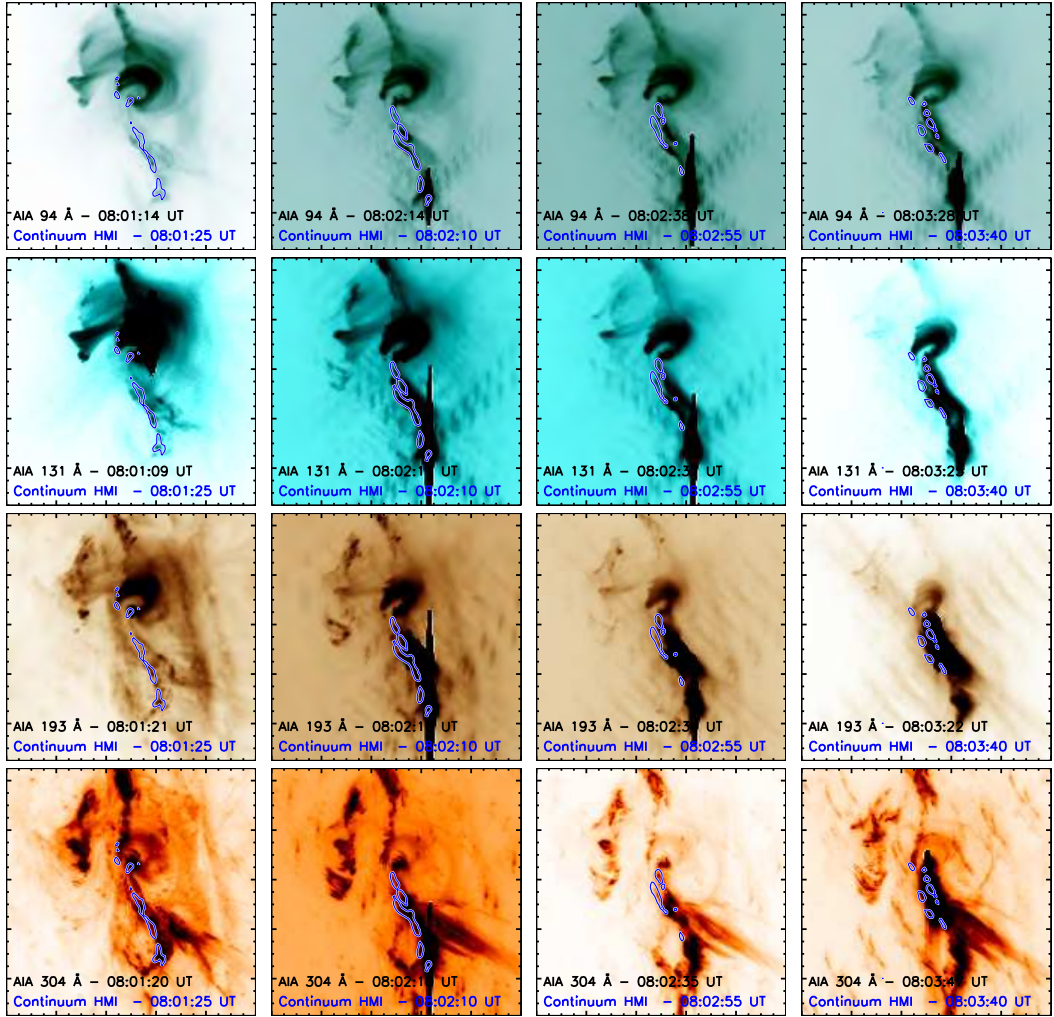

**Figure 4.** Sequence of images in EUV wavebands (in 94, 131, 193 and 304 Å from top-to-bottom rows, respectively) from AIA/SDO during 08:01 - 08:03 UT, corresponding to the peak of the impulsive phase of the flare. Over-plotted are the contours of 30% levels of the maximum of the continuum enhancement obtained from HMI/SDO.

Here $I_b$ represents the background intensity estimated by averaging the values of a rectangle of 30 × 30 pixel area far from the flare affected area. Similarly, $I_f$ represents mean of the counts within the box of 30 × 30 pixels in the flare location. Employing the aforesaid equation, we estimated contrast over three locations (R1, R2 and R3) as shown in Fig. 7. We also show the box corresponding to background location (B1, B2, B3) for each region with the same colour. It has to be noticed that we considered the background region close yet far enough from the respective flare activity box and having the same radial distance in order to account for limb darkening. Top row of Fig. 7 shows the images at 08:08:17 UT and 08:02:10 UT, from Coimbra spectrograph (wavelength ∼6547.55Å) and continuum filtergram from HMI/SDO in the left and right panels, respectively. Spatial intensity contrast estimated from HMI and Coimbra images is shown in the middle and bottom rows, respectively. Middle panel of Fig. 7 shows the temporal evolution of the contrast at aforesaid three locations derived from HMI continuum filtergrams shown by red, blue and green colours, respectively. Bottom panel of Fig. 7 shows the spectral evolution of contrast derived from Coimbra spectrograph observations at 08:08:17 UT.

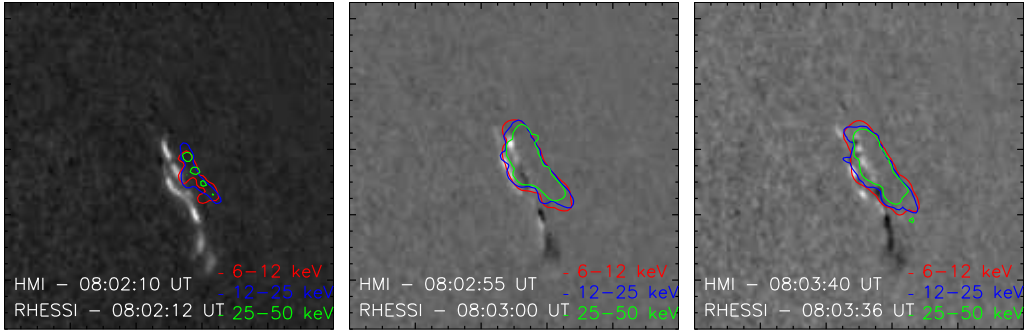

**Figure 5.** Sequence of difference images as derived from continuum images obtained by HMI instrument during the peak emission of the flare. Over-plotted are the co-temporal contours representing 3% of the maximum intensities in 6-12, 12-25 and 25-50 keV drawn in red, blue and green colours, respectively.

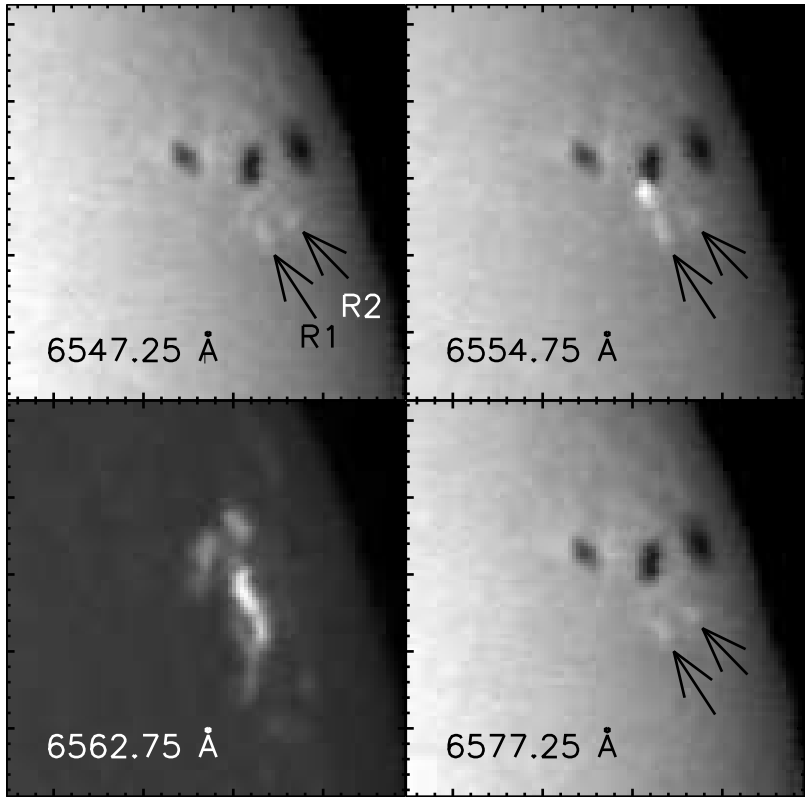

**Figure 6.** Images of the flare region obtained from the Coimbra spectroheliograph at ∼ 08:08:17 UT in 6547.25, 6554.75, 6562.75 and 6577.25 Å. Regions 'R1' and 'R2' are denoted by arrow and can be seen across the images taken in the wings of H$\alpha$ line.

Contrast of the HMI continuum (middle row of Fig. 7) has reached its maximum value of 0.10 at 08:02:55 UT, coinciding with the peak time of the impulsive phase. The intensity contrast corresponding to the far-red and blue wing of the H$\alpha$ line, derived from Coimbra observations at 08:08:17 UT (bottom row), is 0.1, 0.07 and 0.02 for regions R1, R2 and R3, respectively. The contrast values estimated at quasi-continuum wavelengths from Coimbra observations for the same regions match very well with that derived from HMI continuum images at 08:08:17 UT (dotted grey line in the middle row plot). In addition, we may note that the intensity contrast of H$\alpha$ line center is estimated to be $\sim 200\%$. From the Coimbra observations, although R2 appears to be a region of continuum enhancement associated with the flare activity (cf. Fig. 6), temporal evolution of the contrast derived from HMI continuum images for this region shows similar contrast levels even before the flare i.e. at 07:55:00 UT. In this regard, quasi-continuum emission from R2 may also be associated with the plage heating in addition to the flare activity.

## 5. Summary

In this paper we analysed continuum WL emission from the active region. There are two components of the WL continuum emission: from a) plages, and b) flare. The main problem was to disentangle these two components and determine the contribution of both. We have shown that using only solar spectra, without detailed analysis of the time evolution of images, complemented by hard X-ray data, it is difficult to distinguish the continuum emission from flares and faculae. Intensity of both emissions have similar contrast and besides, plage and flare structure are located close together. However, using time series of HMI images we were able to determine the emission separately for both components.

In the next papers we will concentrate on the modeling of the continuum emission from plages and flares. Comparison of the theoretically calculated continuum spectra with observations allows us to determine the physical properties of the emitting plasma and to resolve the ambiguity between plages and flare emission.

ACKNOWLEDGEMENTS. This project has received funding from the European Communitys Seventh Framework Programme (FP7/2007-2013) under grant agreement no. 606862 (F-CHROMA) (AB, AKA, PH). This work is also supported by the grant no. 16-18495S of the Grant Agency of the Czech Republic (AB, PH) and by the institutional project RVO: 67985815 of the Astronomical Institute of the Academy of Sciences of the Czech Republic (AB, PH, MS).

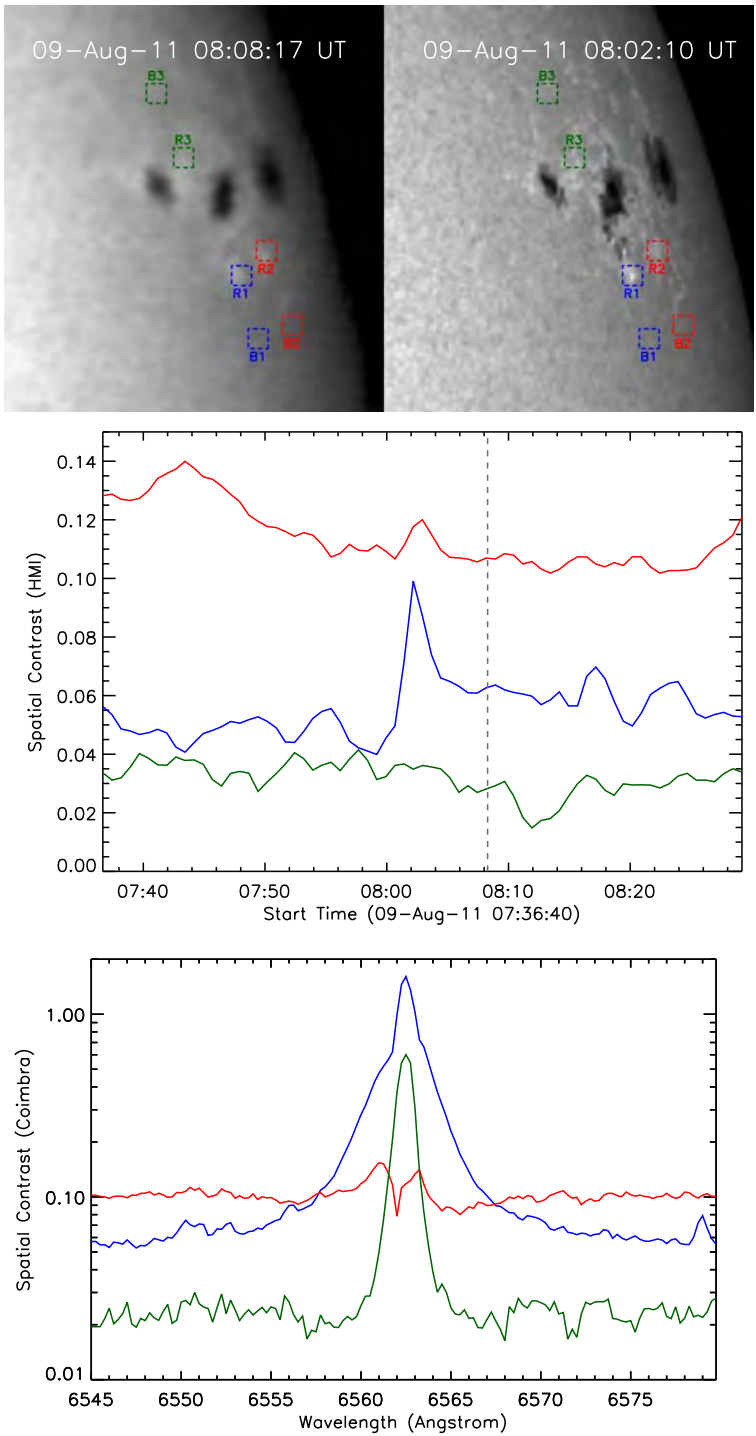

**Figure 7.** Top row: Images at 08:08:17 UT and 08:02:10 UT, from Coimbra spectrograph and HMI/SDO are shown in the left and right panels, respectively. Red blue and green set of boxes represent the location at which the intensity and respective background has been estimated. Middle row: Temporal evolution of the contrast derived from the HMI continuum filtergrams for the aforesaid three location and shown by red, blue and green colors, respectively. Dotted line drawn in grey color represents the time at which Coimbra observations are available. Bottom row: Spectral evolution of contrast derived from Coimbra spectrograph observations at 08:08:17 UT.